\documentclass[twocolumn,showpacs,preprintnumbers,amsmath,amssymb,floatfix,prl]{revtex4}
\usepackage[latin1]{inputenc}
\usepackage{graphicx}
\usepackage{t1enc}
\usepackage[french]{babel}
\usepackage{chngpage}

\begin{document}

\title{Attosecond lighthouses}
\author{H. Vincenti$^{1}$, F. Quéré$^{1,*}$}

\affiliation{CEA, IRAMIS, Service des Photons Atomes et Molécules, F-91191 Gif-sur-Yvette, France}

\begin{abstract}
Coherent light beams composed of ultrashort pulses are now increasingly used in different fields of Science, from time-resolved spectroscopy to plasma physics. Under the effect of even simple optical components, the spatial properties of these beams can vary over the duration of the light pulse \cite{akturk}. In this letter, we show how such spatio-temporally coupled electromagnetic fields can be exploited to produce an attosecond lighthouse, \emph{i.e.} a source emitting a collection of isolated attosecond pulses, propagating in angularly well-separated light beams. This very general effect not only opens the way to a new generation of attosecond light sources, particularly suitable for pump-probe experiments, but also provides a powerful new tool for ultrafast metrology, for instance giving direct access to fluctuations in the phase of the laser field oscillations with respect to the pulse envelop, right at the focus of even the most intense ultrashort laser beams.
\end{abstract}

\maketitle

Ultrashort light beams are said to exhibit spatio-temporal couplings (STC) when their spatial properties depend on time, and conversely \cite{akturk} - i.e. their electric field $E(x,y,z=z_0,t)\neq E_1(t)E_2(x,y)$. The importance of STC has been largely overlooked in most laser-matter interaction experiments, until recently \cite{Vitek}. In strong-field science, STC are even considered as highly detrimental, because they systematically decrease the peak intensity at focus \cite{pretzler}. In this letter, we show that on the opposite, moderate and controlled  STC provide a powerful means of controlling high-intensity laser-matter interactions, and pave the way to a whole range of new experimental capabilities.

To demonstrate this idea, we consider a particular application of STC to attosecond pulse generation (1 $as$=$10^{-18}$ $s$), which has been the key issue in the development of attosecond Science \cite{reviewKrauszIvanov}. All attosecond light sources demonstrated so far are based on high-order harmonic generation (HHG) of intense femtosecond laser pulses in different media \cite{reviewKrauszIvanov,Autoco_CWE}. Since many-cycle long pulses naturally produce trains of attosecond pulses, considerable efforts had to be deployed in the last fifteen years for the development of 'temporal gating' techniques, to isolate single attosecond pulses, more readily usable for time-resolved measurements of electron dynamics in matter. A variety of experimentally-challenging techniques have now been demonstrated for HHG in gases \cite{Sansone, Goulielmakis}.
In contrast, the problem is still unsolved experimentally for HHG on plasma mirrors \cite{2006NJPh....8...19T, Baeva, Naumova}, one of the promising processes to obtain the next generation of attosecond light sources \cite{2009RvMP...81..445T, ThauryJphysB}. We describe here how STC provide a new approach to this problem, of unprecedented simplicity, generality and potential: one of the most basic types of STC, wavefront rotation \cite{akturk} (WFR), can be exploited to generate a collection of single attosecond pulses in angularly well-separated light beams -an attosecond lighthouse- even with relatively long laser pulses.

\begin{figure} [!h]
\includegraphics[width=1\linewidth]{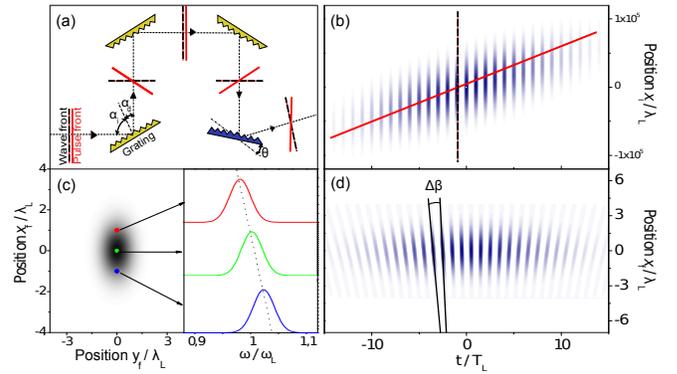}
\caption{\textbf{Pulse front tilt  and wavefront rotation of a femtosecond laser beam.} Panel (a) illustrates how a slightly misalign four-grating compressor leads to a residual pulse front tilt at its output. (b) $E$-field before focusing (real part of Eq. (\ref{E(xi,t)})), exhibiting pulse front tilt.  (c) 2D-Intensity distribution and spectrum at focus. The focal spot is elongated, and the field spectrum varies spatially along the long direction of the ellipse (spatial chirp). (d) Associated $E$-field at focus (real part of Eq. (\ref{E(xf,t)})), exhibiting wavefront rotation in time. Laser pulse of duration before focusing $\tau_i=25$ $fs$, with $w_i=70$ $mm$, $f=200$ $mm$, $\xi=\tau_i/ w_i=0.36$ $fs/mm$, leading to $w_f=2.6 \lambda_L$ and $\tau _f=35$ $fs$ and $v_r=7$ $mrad/fs$. These parameters are typical of a state-of-the-art high-power femtosecond laser.}\label{PFT-WFR}
\end{figure}

Let us first briefly summarize the concept of WFR at the focus of a femtosecond laser beam.  Out of focus, this STC takes a different form, called pulse front tilt \cite{akturk}, which corresponds to an $E$-field of the form (assuming Gaussian profiles in time and space):
\begin{equation}\label{E(xi,t)}
E(x_i,t)=E_0 \exp\left(-2\left[\frac{t-\xi x_i}{\tau_i}\right]^{2}-2\frac{x_i^2}{w_i^2}\right)\times\exp\left(i \omega_L t\right)
\end{equation}
where $x_i$ is one of the transverse spatial coordinate, $\tau_i$ is the Fourier-transform limited pulse duration 
at a given position in the beam, $w_i$ the beam diameter before focusing, $\xi$ the pulse front tilt parameter, and $\omega_L$ the laser central frequency. All widths are defined as full widths at $1/e$ of the intensity profile. Pulse front tilt occurs whenever $\xi\neq 0$, i.e. when the line formed by the pulse maxima in the $(x_i,t)$ space -the pulse front- is tilted with respect to the wavefronts (see Fig. \ref{PFT-WFR}(a) and (b)).

For instance, pulse-front tilt can be induced upon diffraction on a grating \cite{Treacy}: since the incidence angle $\alpha_i$ and diffraction angle $\alpha_d$ are not equal (Fig. \ref{PFT-WFR}(a)), the accumulated optical path length varies with transverse coordinate $x_i$, resulting in a varying delay between the wavefront and pulse front across the beam. In Chirped-Pulse Amplification lasers, residual pulse front tilt thus occurs whenever some of the gratings used in the compressor are not exactly parallel (see Fig. \ref{PFT-WFR}(a)). 

This example provides a straightforward way of understanding the $E$-field configuration once a beam with initial pulse front tilt is focused.  Indeed, a grating followed by a focusing element acts as spectrometer. At focus, such a beam thus presents a spectrum which depends on space (see Supplementary Information), as illustrated in Fig. \ref{PFT-WFR}(c). This effect is known as spatial chirp in the $(x_f,\omega)$ space , where $x_f$ is the transverse spatial coordinate at focus, and $\omega$ the light frequency. As $\xi$ increases, the focal spot becomes more and more elliptical, with a long axis along the direction of spatial chirp (see Fig. \ref{PFT-WFR}(c)).

In the $(x_f,t)$ domain, such a spatial chirp implies that the time spacing between successive wavefronts varies with the transverse coordinate $x_f$: as a result, the wavefronts rotate in time (see Fig. \ref{PFT-WFR}(d)). Indeed, Fourier-transforming Eq.(\ref{E(xi,t)}) with respect to $x_i$ leads to the field $\tilde{E}(x_f,t)$ at the focus of an optics of focal length $f$:
\begin{eqnarray}\label{E(xf,t)}
\tilde{E}(x_{f},t)&\propto& \exp \left(-2\frac{t^{2}}{\tau_{f}^{2}}-2\frac{x_{f}^2}{w_{f}^{2}}\right)\times\exp(i\varphi(x_f,t)) \nonumber\\ 
\varphi(x_f,t)&=&4\frac{w_{i}\xi}{w_{f}\tau_{f}\tau_{i}}x_{f}t+\omega_{L}t 
\end{eqnarray}
where the pulse duration at focus, $\tau_f$, and the effective beam waist along the $x_f$ axis, $w_f$, are given by
\begin{eqnarray}\label{taufwf}
\tau_f=\tau_i\sqrt{1+(w_i \xi/\tau_i)^2} \nonumber\\ 
w_f=w_0 \sqrt{1+(w_i \xi /\tau_i)^2}
\end{eqnarray}
with $w_0=2(\lambda f/\pi w_i)$ the usual beam waist at focus when $\xi=0$. The instantaneous direction of propagation of light $\beta$ is given by $\beta \approx k_\bot/k_L$, where $k_L=\omega_L/c$ is the laser wave vector, and $k_\bot=\partial \varphi /\partial x_f$ its transverse component. The $x_f t$ term in $\varphi$ implies that $\beta$, and hence the instantaneous wavefront direction, vary in time, with a rotation velocity $v_r=d\beta/dt$ deduced from Eq. (\ref{E(xf,t)}):
\begin{equation}
v_r=\frac{w_i^2}{f\tau_i^2}\frac{\xi}{1+\left(w_i \xi/\tau_i\right)^{2}}
\end{equation}
For given duration $\tau_i$ and numerical aperture $\theta_L=w_i/f$ of the beam, $v_r$ reaches its maximum possible value $v_r^{max}=w_i/2f \tau_i=\theta_L/2\tau_i$ for $\xi_0=\tau_i/w_i$. Intuitively, this maximum is achieved when light roughly sweeps the angular width $\theta_L$ of the beam, in the shortest achievable duration for this pulse. This optimum case corresponds to a pulse front delay of $\tau_i$ across the laser beam diameter $w_i$ before focusing, while at focus $w_f=\sqrt{2}w_0$ and $\tau_f=\sqrt{2}\tau_i$, leading to a reduction of the peak intensity by a factor of 2 only compared to the case when $\xi=0$. For a 25 fs laser pulse focused with $\theta_L=0.35$ $rad$ ($w_i \approx f/3$), $v_r^{max}$ reaches the considerable value of $7$ $mrad/fs$, i.e. $\approx10^{12}$ revolutions per second. 

\begin{figure*} 
\centering
\includegraphics[width=0.9\linewidth] {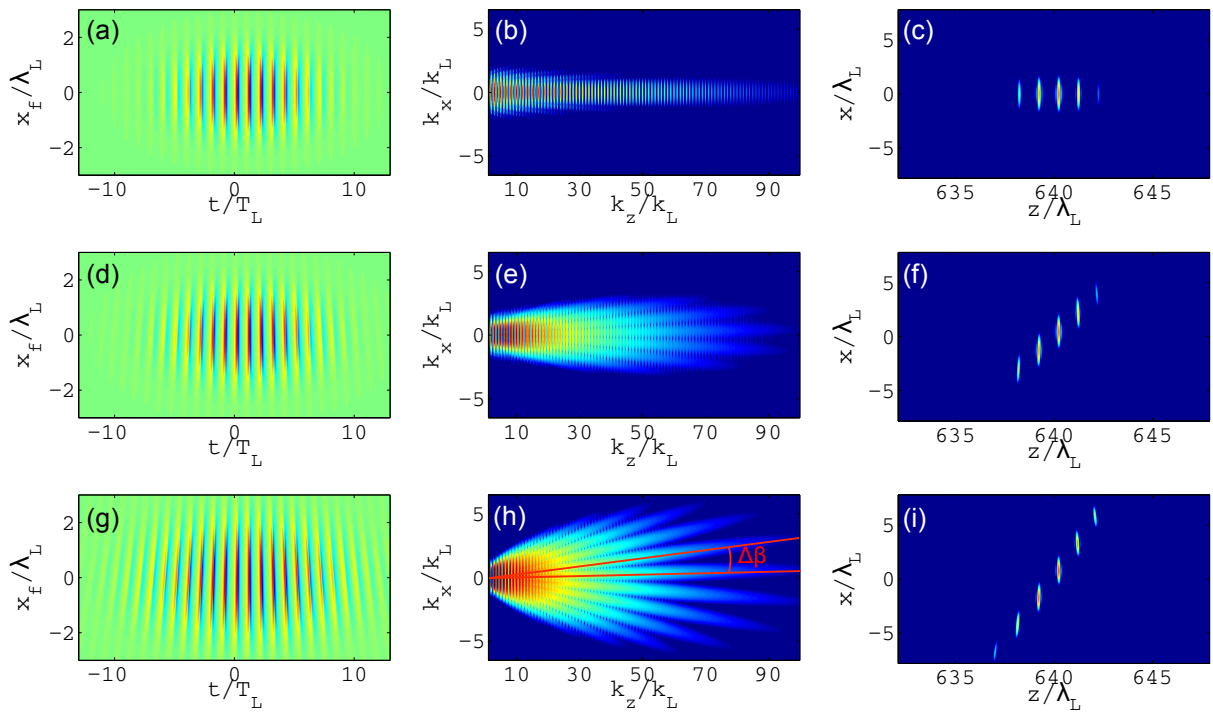}
\caption{ \textbf{The attosecond lighthouse effect}. The left panels display the electric field after the HHG interaction in the $(x_f,t)$ space, provided by the ROM model. The parameters of the laser beam are the same as in Fig. \ref{PFT-WFR}, and the normalized vector potential $a_0=eE/mc\omega_L=3$ ($e$ and $m$ electron mass and charge, $c$ light velocity, and $E$ laser field amplitude). The temporal distortion of the field induced by HHG is clearly visible. The middle panels show the 2D Fourier transforms of these fields. In these plots, the distance from origin corresponds to frequency, and the polar angle corresponds to the propagation direction. The right panels show instantaneous snapshots of the intensity of the attosecond pulses corresponding to harmonic orders 70 to 80, as a function of transverse and longitudinal spatial coordinates $x$ and $z$, at a distance from the source larger than the Rayleigh length of the corresponding harmonics. Note that assuming constant laser intensity, the \emph{peak} spectral signal decreases by a factor $N^2$ between panels ((b) and (h) (while the total integrated signal increases), due to (1) the dispersion of the $N$ attosecond pulses in different directions, and (2) the resulting absence of constructive interferences at harmonic frequencies of the laser. }\label{Model}
\end{figure*}
	
WFR has an immediate and far-reaching interest in the context of attosecond pulse generation. When a usual laser pulse is used to drive HHG in gases or plasmas, the generated train of $N$ attosecond pulses is generally produced within a collimated beam, with a divergence smaller than that of the initial laser beam. In contrast, in the presence of WFR, due to the coherence of HHG, these $N$ attosecond pulses are all emitted in slightly different directions, which correspond to the instantaneous directions of propagation of the laser field at the times of generation. 

This attosecond lighthouse effect can be exploited to produce $N$ angularly well-separated short-wavelength beams, each one containing a single attosecond pulse. 
 To achieve such a result, the rotation $\Delta \beta=v_r \Delta t$ of the wavefront in the time interval $\Delta t$ between the emissions of two successive attosecond pulses has to be larger than the divergence $\theta_n$ of the short-wavelength light beam around frequency $n \omega_L$. Since the maximum achievable value of $v_r$ is $v_r^{max}=\theta_L/2\tau_i$, this is only possible provided the following general condition is fulfilled:
\begin{equation} \label{resolution}
\frac{\theta_n}{\theta_L} \leq \frac{1}{\alpha p N_c}
\end{equation}
where $N_c \geq N$ is the number of optical cycles in the driving laser pulse, $p$ the number of attosecond pulses generated every laser optical cycle, and $\alpha=O(1)$ a prefactor which depends on the intensity contrast required between a given attosecond pulse and its first satellites (see Supplementary Information). The key issue to exploit this effect is thus to achieve a sufficiently small divergence of the harmonic beam, which critical value increases as the duration of the driving laser pulse decreases, according to Eq.(\ref{resolution}).

To validate the basic physics previously described, we now focus on the case of HHG on plasma mirrors in the relativistic interaction regime. As an intense laser field reflects on a dense plasma with a sharp surface, it induces an oscillation of the plasma surface, which generates through the Doppler effect high-order harmonics in the reflected beam, associated to trains of attosecond pulses. We first describe this process with the simple Relativistic Oscillating Mirror (ROM) model described in ref. \cite{Lichters}, which we use to calculate the resulting $E-$field (see the Methods section), in the presence of WFR. 

The predictions of this model are displayed in Fig. \ref{Model}, for different WFR velocities, from $0$ to $v_r^{max}$. 
As expected, in the absence of WFR, harmonics of the laser frequency are emitted in a single collimated beam (Fig. \ref{Model}(b)), with a divergence weaker than that of the fundamental frequency, and are associated in the time domain to a train of $5$ attosecond pulses (Fig. \ref{Model}(c)). As $v_r$ increases, this single beam progressively splits into a set of different beams (Fig. \ref{Model}(e,h)). At the optimum rotation velocity, these beams are angularly well-separated (Fig. \ref{Model}(h)), each of them carries a continuous electromagnetic spectrum, and is associated to a single attosecond pulse (Fig. \ref{Model}(i)).

To confirm the predictions of this simple model, we have performed 2D Particle-In-Cell (PIC) simulations of HHG on plasma mirrors, in the ROM regime \cite{ThauryJphysB} ($a_0=6$), including WFR (see the Methods section). 
The results of such simulations are displayed in Fig. \ref{PIC}(a) and (b), with and without WFR. The attosecond lighthouse effect is clearly observed (see also movie in Supplementary Information), and the rotation velocity is large enough, and the short-wavelength light beam divergence small enough, to isolate a single attosecond pulse by simply setting a slit in the far field (Fig. \ref{PIC}(b)). This shows that fulfilling Eq.(\ref{resolution}) is physically realistic for plasma mirrors.

\begin{figure}
\includegraphics[width=\linewidth] {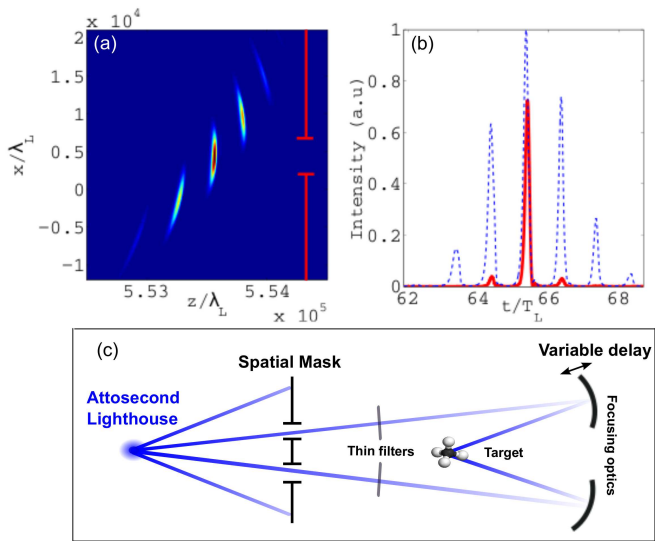}
\caption{\label{PIC}  \textbf{Particle-in-Cell simulation of the attosecond lighthouse effect}. The left panels shows the far-field intensity distribution of the attosecond pulses (harmonic orders 15 to 30) generated when an intense laser field ($a_0=6$, $w_f=8 \lambda_L$, $\tau_f=22$ $fs$, $\theta_L=56$ $mrad$), with a WFR velocity $v_r=v_r^{max}=1.3$ $mrad/fs$, reflects on a plasma mirror where it induces the ROM effect. These physical parameters are experimentally realistic, and the obtained ratio of harmonic and laser divergences consistent with experimental data \cite{2009NatPh...5..146D}. The right panel shows the angularly-integrated temporal intensity profile of this field (red curve), after filtering by the diaphragm displayed in red on the left panel, which selects $\approx 60$ \% of the main pulse energy. The blue curve corresponds to the temporal profile obtained from a PIC simulation in the same interaction conditions, but without WFR. Both curves have been normalized by the peak intensity of the train without WFR. Panel (c) shows a schematic layout of an attosecond pump-probe experiment that uses a lighthouse source.}
\end{figure}

Since the attosecond lighthouse effect relies on the universal relationship between the laser phase $\varphi$ and the harmonic phases $\varphi_n$, $\varphi_n=n\varphi$, it in principle applies to any HHG mechanism. It will however be particularly relevant in the case of plasma mirrors, since its implementation only requires a slight rotation of one of the gratings of the compressor, or introducing a prism in the beam (see Supplementary Information). This is by no means comparable to the experimental complexity of the gating techniques proposed so far for plasma mirrors \cite{2006NJPh....8...19T, Baeva, Naumova}.

Compared to all other gating methods in any generation medium, the attosecond lighthouse effect also provides an ideal optical scheme for attosecond pump-probe experiments, which avoids critical temporal jitter issues related to the use of multiple laser pulses. Indeed, two or more \emph{perfectly synchronized} single attosecond pulses can be generated \emph{with a single laser pulse}, in spatially separated beams, by setting adequate spatial masks in the far field (Fig. \ref{PIC}(c)). These multiple pulses can then be manipulated with a few independent optics, and recombined with a variable delay on a target.

This effect also has great potential in terms of metrology. First, it provides a direct way to study the properties of all individual attosecond pulses generated along a laser pulse. By measurements on the different beams,  the spectrum, divergence, relative energy of each of these individual pulses can be determined. This will particularly useful in experiments that exploit HHG as a probe of the generating medium, since it provides a stroboscope that for instance registers the temporal evolution of molecular orbitals \cite{2004Natur.432..867I, 2010Natur.466..604W} with attosecond resolution over the whole laser pulse duration , without the need to scan any delay.
Second, with WFR, the propagation directions of the $N$ attosecond pulses depend on their times of emission, and thus provide direct information on the physics of the generation process. Besides, since these emission times depend linearly on the Carrier-Envelop relative Phase (CEP) \cite{CEP} of the driving laser pulse, changes in the CEP result in shifts of the emission angular pattern under the beam spatial envelop, as illustrated in Fig. \ref{CEP}. On the one hand, this implies that CEP stabilization is still required to achieve shot-to-shot reproducibility of this source. But on the other hand, when CEP is not stabilized, measuring the harmonic beam angular pattern provides a straightforward means of tracking CEP variations right at focus, and of binning any experimental data obtained on different laser shots as a function of its actual value. This for instance opens the way to CEP-resolved experiments in the relativistic interaction regime.

\begin{figure}
\includegraphics[width=0.85\linewidth] {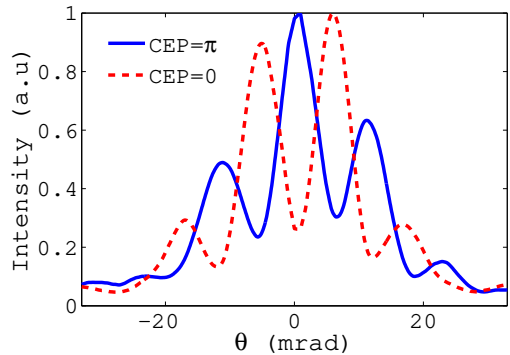}
\caption{ \label{CEP}  \textbf{Sensitivity to the Carrier-Envelop relative Phase of the driving laser}. The curves display the spatial profile of the spectrally-integrated harmonic signal (from orders 25 to 30) in the far-field, for two values of the driving-laser pulse CEP differing by $\pi$, obtained from PIC simulations, for the same physical parameters as in Fig. \ref{PIC}.}
\end{figure}

In conclusion, wavefront rotation constitutes a very powerful tool for attosecond Science, and provides an ideal scheme to isolate highly-collimated attosecond pulses in the X-ray range  from beams reflected on plasma mirrors \cite{2004PhRvL..93k5002G, dromey_NP, dromey_prl}. More generally, while spatio-temporal couplings of ultrafast laser beams have so far mostly been considered as detrimental, this paper illustrates how shaping light fields in both time and space can provide new degrees of freedom to manipulate matter with intense light, leading to new experimental capabilities.




\begin{center}
\textbf{Methods}
\end{center}

\textbf{The oscillating mirror model:} The model we use is similar to the one developed in ref.\cite{Lichters}, adapted to a 2D case with WFR. The position $Z(x_{f},t)$ of the mirror surface at the transverse position $x_f$ follows an harmonic oscillation at the laser frequency, with peak velocity $v_{m}(x_f,t)/c=a/\sqrt{1+a^2}$, where $a(x_f,t)$ is the spatio-temporal envelope of the incident laser. This leads to $Z(x_f,t)=(v_{m}/\omega_{L})\cos \varphi(x_f,t)$, with $\varphi$ given by Eq.(\ref{E(xf,t)}).  The reflected field $E_{r}(x_f,t)$ is then proportional to  $\tilde{E}(x_{f},t_{ret})e^{ik_LZ(t_{ret})}$, where $t_{ret}(x_f)$ is the solution of $Z(x_f,t_{ret})=c(t-t_{ret})$. 

\textbf{Particle-In-Cell (PIC) simulations:} We use the PIC code CALDER in 2D. The plasma has a maximum density of $100n_c$, an initial density gradient of $\lambda_L/200$, an initial electronic temperature of 0.1 $keV$, and ions are mobile. The incidence angle is 45$\degres$, and the laser field is $p$-polarized. This field is injected in the simulation box through boundary conditions, in the form of Eq. (\ref{E(xf,t)}).  The size of the simulation box is $40\lambda_L\times 75\lambda_L$, with a mesh size of $2.9\times10^{-3}\lambda_L$, the time step is $2\times10^{-3}T_L$, and we use 20 particles/cell. A typical calculation requires 48 hours on 512 CPUs.

\textbf{Propagation of the reflected field:} Once the reflected field $E_{r}(x_f,z=0,t)$ right after the target is obtained (either from the model or from PIC simulations), we calculate its propagation in vacuum using plane waves decomposition (i.e. by applying the phase term $\exp(ic k_y^2z/2\omega )$ to the 2D Fourier-Transforms $\hat{E}(k_y,\omega)$ of Fig. \ref{Model}(b-e-h), and then calculating the 2D inverse Fourier-transform), in order to determine its structure $E_{r}(x,z,t)$ at arbitrary distance $z$ from the target.


\begin{center}
\textbf{\large Supplementary Information for "Attosecond Lighthouses"}
\end{center}

\renewcommand{\figurename}{Figure S -}
\maketitle

This supplementary information contains:
\begin{enumerate}
\item a calculation of the spatially-resolved laser spectrum at focus in the presence of WFR, 
\item a derivation of the pulse front tilt $\xi[fs/mm]$ caused by a single prism or a misaligned four-grating compressor,
\item a calculation of the maximum acceptable ratio between harmonic and laser divergences to obtain an isolated attosecond pulse, as a function of the required contrast with its sattelites, and of the percentage of the main pulse energy selected by a diaphragm. \\
\end{enumerate}

\textbf{\normalsize 1. Calculation of the laser spectrum at focus in the presence of WFR}\\

The spectrum $\hat{E}(x_{f},\omega)$ of the  fied $\tilde{E}(x_{f},t)$ at focus is given by the following integral:

\begin{equation*}
\hat{E}(x_{f},\omega)\propto\int_{-\infty}^{\infty}\tilde{E}(x_{f},t)e^{i\omega t}dt
\end{equation*}

This yields:  

\begin{equation}
\hat{E}(x_{f},\omega)\propto \exp\left[-\left(\left(\omega-\omega_{L}\right)+4\frac{\xi w_{i}}{\tau_{i}\tau_{f}}\frac{x_{f}}{w_{f}}\right)^{2}\frac{\tau_{f}^2}{8}-2\frac{x_{f}^2}{w_{f}^2}\right]\end{equation}

The above equation shows the effect of spatial chirp at focus, \emph{i.e.} the fact that the central frequency of the spectrum varies with $x_f$.\\

\textbf{\normalsize 2. Pulse-front tilt calculations}\\

\textbf{ \small 2.1 Relation between angular dispersion and pulse front tilt}\\

The reference by J.Hebling, \textit{Optical and Quantum Electronics 28(1759-1763) July(1996)}  demonstrates that pulse-front tilt and angular dispersion introduced by a dispersive element are related through the following relation:
\begin{equation}
\tan c\xi=\overline{\lambda}\dfrac{d\Gamma}{d\lambda}
\label{psftang}
\end{equation}
where:\\
\begin{itemize}
\item $\xi$ denotes the pulse front tilt,
\item $\Gamma(\lambda)$ is the angle of emergence from the dispersive element of light at wavelength $\lambda$, and $\dfrac{d\Gamma}{d\lambda}$ is thus the angular dispersion.
\item $\overline{\lambda}$ is the mean wavelength of the the incident light on the dispersive element. \\
\end{itemize}
Using Eq.(\ref{psftang}), we next derive the pulse front tilt caused by a single prism, and the misalignment of a four-grating optical compressor. \\

\textbf{\small 2.2 Derivation of the pulse-front tilt $\xi[fs/mm]$ caused by a  single prism}\\

In the following,  we consider a prism with an apex angle $\beta$ and a refractive index $n(\lambda)$ given by the Sellmeier relation: 

\begin{equation}
n(\lambda)=\sqrt{1+\dfrac{B_{1}\lambda^{2}}{\lambda^2-C_{1}}+\dfrac{B_{2}\lambda^2}{\lambda^{2}-C_{2}}+\dfrac{B_{3}\lambda^2}{\lambda^{2}-C_{3}}}
\label{Sellmeier}
\end{equation}

where, in the case of fused silica:

\begin{center}
{\renewcommand{\arraystretch}{1} 
{\setlength{\tabcolsep}{0.05cm} 
\begin{tabular}{|c|c|c|}
\hline
$B_{1}$&$B_{2}$&$B_{3}$\\
\hline
$1.03961212$&$0.231792344$&$1.01046945$\\
\hline
$C_{1}[\mu m^2]$&$C_{2}[\mu m^2]$&$C_{3}[\mu m^2]$\\
\hline
$6.00069867\times10^{-3}$&$2.00179144\times10^{-2}$&$1.03560653\times 10^{2}$\\
\hline
\end{tabular}}}
\end{center}

\begin{figure} 
\centering
\includegraphics[width=0.6\linewidth]{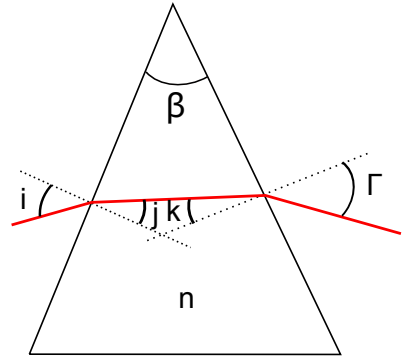}
\caption{\footnotesize Prism with a refractive index $n$ and  an apex angle  $\beta$. The red line shows a trajectory of a light ray passing through the prism.  }
\label{prisme}
\end{figure}

Let's first calculate the angle  $\Gamma$  of emergence of an incident ray - with incidence angle $i$ -  from the normal to the  second face of this prism (cf. Fig. S - \ref{prisme}).

Snell-Descartes laws of refraction  give :

\begin{eqnarray}
\sin i&=&n \sin j\\
 \sin \Gamma&=&n \sin k
\end{eqnarray}

Moreover , the angles $j$, $k$ and  $\beta$ satisfy the following  geometrical relation:

\begin{equation}
\beta=j+k
\end{equation}

It yields:

\begin{equation}
\Gamma=\arcsin\left[n(\lambda)\sin\left(\beta-\arcsin\frac{\sin i}{n(\lambda)}\right)\right]
\label{eqprisme}
\end{equation}

For simplicity, we only consider the case $i=0$, which leads to: 

\begin{equation} \label{psft1}
\tan c\xi=\overline{\lambda}\dfrac{d\Gamma}{d\lambda}=-\overline{\lambda}\dfrac{dn}{d\lambda}\dfrac{\sin\beta}{\sqrt{1-n^2\sin^2\beta}}
\end{equation}

From Eq.(\ref{psft1}), the apex angle $\beta$ required to get a pulse front tilt $\xi=\tau_{i}/w_{i}=0.5fs/mm$, that maximizes the wavefront rotation after focusing of a laser beam with $\tau_{i}=25fs$ and $w_{i}=50mm$, is $\beta=0.5\degres$.\\

\textbf{\small 2.3 Derivation of the pulse-front tilt $\xi[fs/mm]$ caused by a misaligned four-grating optical compressor}\\

Misaligning the last grating  of a four-grating compressor (represented on Fig. 1(a) of the Letter)  by an angle $\theta$ induces a residual pulse front tilt $\xi$, which we calculate in this section.

 We assume that all gratings are identical and that the two first gratings are perfectly aligned, so they do not generate any pulse front tilt.

\begin{figure}[h!]
\centering
\includegraphics[width=0.6\linewidth]{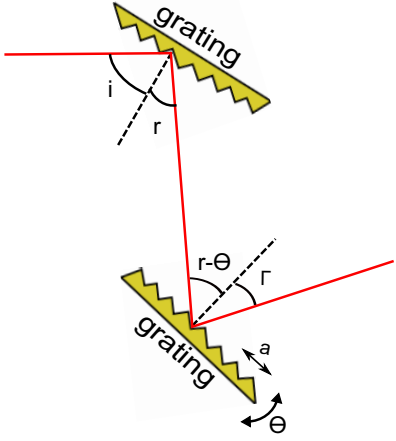}
\caption{\footnotesize Misalignment  of the second pair of gratings of the compressor by an angle $\theta$.  }
\label{reseau}
\end{figure}

In order to calculate  $\xi$, we need to calculate, as in section 2,  the angular dispersion $d\Gamma/d\lambda$ - where $\Gamma$ is the angle represented on Fig. S - \ref{reseau} - created by the misalignement of the last pair of gratings .

By working at the first diffraction order for both gratings we get:

\begin{eqnarray}
\sin i+\sin r&=&\dfrac{\lambda}{a}\\
\sin (r-\theta)+\sin\Gamma&=&\dfrac{\lambda}{a}
\label{reseau2}
\end{eqnarray}

\begin{itemize}
\item where $i$, $r$, $\theta$ and $\Gamma$ are the angles represented on Fig. S - \ref{reseau},
\item $a$ is the grating period. \\
\end{itemize}

Differentiating Eq.(\ref{reseau2}) yields the following equations: 

\begin{eqnarray}
\dfrac{dr}{d\lambda}&=&\dfrac{1}{a\cos r}\\
\dfrac{d\Gamma}{d\lambda}\cos\Gamma&=&\dfrac{1}{a}-\cos(r-\theta)\dfrac{dr}{d\lambda}
\end{eqnarray}

i.e

\begin{equation}
\dfrac{d\Gamma}{d\lambda}=\dfrac{1}{a\cos\Gamma}\left(1-\dfrac{\cos(r-\theta)}{\cos r}\right)
\end{equation}

Hence:

\begin{equation}
\dfrac{d\Gamma}{d\lambda}=\dfrac{1-\dfrac{\cos(r-\theta)}{\cos r}}{a\sqrt{1-\left[\frac{\lambda}{a}-\sin(r-\theta)\right]^2}}
\end{equation}

Assuming that $\theta$ is a small angle, we get:

\begin{equation}
\tan c\xi\approx-\dfrac{\overline{\lambda}}{a}\dfrac{\tan r}{\cos i}\theta
\label{formule4reseaux}
\end{equation}

\textbf{\small Numerical application}\\

From Eq.(\ref{formule4reseaux}), the  misalignement angle $\theta$ required to obtain a pulse front tilt  that maximizes the wavefront rotation velocity at focus is: \\
\begin{enumerate}
\item  is $\theta\approx0.017°\approx1'$ in the case of laser beams used for HHG on plasma mirrors ($\tau_{i}=25fs$ and $w_{i}=50mm$).
\item  is $\theta\approx5'$ in the case of  laser beam used for HHG in gases $\tau_{i}=25fs$ and $w_{i}=10mm$ . \\
\end{enumerate}
Most CPA lasers however do not use four-grating compressor, but rather compressors with two gratings only, used in a double-pass configuration. We have checked that the pulse front tilt induced in such a compressor by a small rotation of one of the gratings is to a very good approximation twice the value provided by Eq.(\ref{formule4reseaux}).\\

\textbf{\normalsize 3. Maximum acceptable ratio between harmonic and laser divergences to obtain an isolated attosecond pulse}\\

\textbf{\small 3.1 Problematics}\\

\begin{figure}[h!]
\centering
\includegraphics[width=\linewidth]{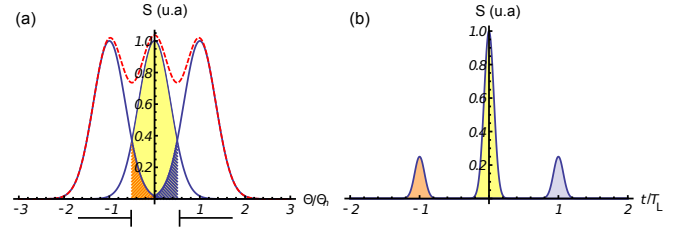}
\caption{\footnotesize (a) Angular profiles of  three  attosecond pulses  generated by the attosecond lighthouse effect.  The blue Gaussians correspond to the angular profiles of the single attosecond pulses. The red dashed curve, which corresponds to the sum of the three blue curves, is the signal which would be recorded on a detector placed on a screen very distant from the source. By setting a slit in the beam, we select the yellow part of the central pulse as well as the edges of the satellite beams (blue and orange shaded portions). (b) Angularly-integrated temporal intensity profile of the light beam selected by the slit shown in (a). The yellow central pulse corresponds to the yellow area in (a), while its satellites in  orange/blue  correspond to the orange/blue shaded areas in (a). Here, the intensity contrast between the satellites and the central pulse is $\gamma=0.1$.  }
\label{traceThetaT}
\end{figure}

In the attosecond lightouse method, an isolated attosecond pulse can be produced by using a slit to spatially filter, in the far field, the electromagnetic field generated by the target. The choice of the width of this slit is the result of a compromise between two contraints (see Fig. S - \ref{traceThetaT}(a)):
\begin{enumerate}
\item The smaller the width of the slit, the higher the contrast ratio $1/\gamma$ between the energy of the main attosecond pulse and its satellites at $\pm T_L$, coming from the adjacent beams.
\item The larger the width of the slit, the larger the fraction $f$ of the total energy of the main attosecond beam that goes through this slit.
\end{enumerate}
The wavefront rotation velocity defines the angular shift $\Delta \beta$ between adjacent attosecond beams, through $\Delta \beta=v_r \Delta t$, where $\Delta t=T_L/p$ is the time interval between the emission of two successive attosecond pulses ($p$ is the number of attosecond pulses generated per optical period). For given laser parameters, the maximum possible value of $\Delta \beta$ is determined by the maximum value $v_r^{max}=\theta_L/2\tau_i$ of $v_r$, and is given by $\Delta \beta ^{max}=\theta_L/2pN_c$, where $N_c$ is the number of optical cycle in the laser pulse before focusing. In the following, we only consider this optimal case, which can easily be achieved experimentally.

Once at maximum $\Delta \beta$, when a slit of given width is set into the beam, the obtained contrast ratio $\gamma$ and energy fraction $f$ depend on the divergence of the harmonic beam $\theta_n$ (see Fig.  S -\ref{traceThetaT}): the larger the harmonic divergence, the lower the contrast ratio, and the smaller the energy fraction $f$. In this section, we calculate the minimum acceptable ratio  $\theta_L/\theta_n$ that makes it possible to obtain given values of $\gamma$ and $f$. \\

\textbf{\small 3.2 Derivation of the critical divergence}\\

Let's note $E_{1}$ the energy of the main attosecond pulse after the slit (yellow area in Fig.  S -\ref{traceThetaT}), and $E_{2}$ the energy of one of the satellites (blue or orange shaded area in Fig. S -\ref{traceThetaT}(a), assumed to be identical). We  assume that all attosecond pulses are Gaussian, with the same total energies $E_T$ and full width at $1/e$, $\theta_{n}$. The main attosecond beam is supposed to be centered on $\theta=0$, so that its satellites are then centered at $\pm\Delta \beta^{max}$.

\begin{figure}[h!]
\centering
\includegraphics[width=\linewidth]{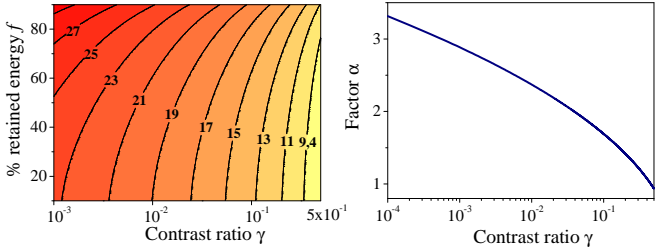}
\caption{\footnotesize (a) Required value of $\theta_L/\theta_n$ to obtain given values of $\gamma$ and $f$, in the case of a laser pulse with $N_c=9$ optical periods, and $p=1$. (b) Factor $\alpha$ defined by Eq.(5) of the letter, as a function of the required contrast $\gamma$, for $f=60$ \%.}
\label{gamvsalph}
\end{figure}

The contrast ratio $\gamma=E_2/E_1$, where $E_1$ and $E_2$ are given by
\begin{eqnarray}
E_1&=& f E_T\\
E_{2}&=&\frac{E_T}{\sqrt{\pi}}\int_{-2\frac{\Delta \theta+\Delta\beta^{max}}{\theta_n}}^{2\frac{\Delta \theta-\Delta\beta^{max}}{\theta_n}}e^{-\Gamma^{2}} d\Gamma
\end{eqnarray}
where we made the change of variable $\Gamma=2(\theta-\Delta\beta^{max})/\theta_{n}$. Here, the angle $\Delta \theta$ corresponds to half the angular width of the slit. This angle can be determined as a function of the retained energy fraction $f$ of the main beam, using:
\begin{eqnarray}
f=\frac{\int_{-\Delta \theta}^{\Delta\theta}e^{-4\theta^{2}/\theta_n^2}d\theta}{\int_{-\infty}^{\infty}e^{-4\theta^{2}/\theta_n^2}d\theta}
\end{eqnarray}

Using the Gauss error function 
\begin{equation}
\chi(z)=\dfrac{2}{\sqrt{\pi}}\int_{0}^{z}e^{-\zeta^2}d\zeta
\end{equation}
and its inverse function $\chi^{-1}$, we obtain:
\begin{equation}
\Delta \theta= \dfrac{\theta_n}{2} \chi^{-1}(f)
\end{equation}

This leads to the following expression for $\gamma$: 

\begin{eqnarray*}
\gamma&=&\dfrac{E_{2}}{E_{1}}\\
       &=&\dfrac{\chi(2\Delta \beta ^{max}/\theta_{n}+\chi^{-1}(f))-\chi(2\Delta \beta ^{max}/\theta_{n}-\chi^{-1}(f))}{2f}
\end{eqnarray*}

This equations determines $\gamma$ as a function of $\theta_n$ and $f$. We used Mathematica to invert it and determine the value of $\theta_L/\theta_n$ required to reach given values of $\gamma$ and $f$. The result of this calculation is plotted in Fig.  S -\ref{gamvsalph}(a), for a pulse with $N_c=9$ optical periods and for $p=1$ (corresponding to HHG on plasma mirrors).

The numbers displayed on this graph for $\theta_L/\theta_n$ can be interpreted in a different way. Indeed, physically, the smallest divergence that can be achieved for a given harmonic order is $\theta_n=\theta_L/n$, when the harmonic source has a size matching the laser spot size and a flat spatial phase. Therefore, the displayed numbers can alternatively be interpreted as the \emph{minimum possible} harmonic order, beyond which given values of $f$ and $\gamma$ can be obtained by the attosecond lighthouse effect.

Finally, we relate this result to the $\alpha$ parameter that appears in Eq.(5) of the Letter. The value of $\alpha$, required  to reach given values of $f$ and $\gamma$, can be deduced from a graph such as the one in Fig. S-\ref{gamvsalph}(a), through the equation $\alpha=\theta_L/\theta_npN_c$. The curve in Fig. S - \ref{gamvsalph}(b) thus shows the evolution of $\alpha$, for a retained energy fraction $f=60$ \%, as a function of the required contrast $\gamma$.

\end{document}